\documentclass[%
reprint,
superscriptaddress,
 amsmath,amssymb,
 aps,
 pre,
]{revtex4-2}
\usepackage[utf8]{inputenc}
\usepackage[T1]{fontenc}
\usepackage{graphicx}
\usepackage{float}
\usepackage{color}
\usepackage{xr}
\usepackage{bm}
\usepackage{rotating}
\usepackage[left=1.5cm, right=1.5cm, top=1.785cm, bottom=2.0cm]{geometry}
\begin{document}
	\title{Novel turbulence and coarsening arrest in active-scalar fluids}
 %\\ Or\\Turbulence in viscous binary fluid mixtures}	
	\author{Nadia Bihari Padhan}
	\email[]{nadia@iisc.ac.in}
		\affiliation{Centre for Condensed Matter Theory, Department of Physics, Indian Institute of Science, Bangalore, 560012, India.}
    \author{Kolluru Venkata Kiran}
	 \email[]{kollurukiran@iisc.ac.in}
	\affiliation{Centre for Condensed Matter Theory, Department of Physics, Indian Institute of Science, Bangalore, 560012, India.}
	\author{Rahul Pandit}
		\email[]{rahul@iisc.ac.in}
	\affiliation{Centre for Condensed Matter Theory, Department of Physics, Indian Institute of Science, Bangalore, 560012, India.}

\date{\today}

\begin{abstract}
We uncover a new type of turbulence -- activity-induced homogeneous and isotropic turbulence in a model that has been employed to investigate motility-induced phase separation (MIPS) in a system of microswimmers. The active Cahn-Hilliard-Navier-Stokes equations (CHNS), also called active model H, provides a natural theoretical framework for our study. In this CHNS model, a single scalar order parameter $\phi$, positive (negative) in regions of high (low) microswimmer density, is coupled with the velocity field $\bm u$. The activity of the microswimmers is governed by an activity parameter $\zeta$ that is positive for \textit{extensile} swimmers and negative for \textit{contractile} swimmers.  With extensile swimmers, this system undergoes complete phase separation, which is similar to that in binary-fluid mixtures. By carrying out pseudospectral direct numerical simulations (DNSs), we show, for the first time, that this model (a) develops an emergent nonequilibrium, but statistically steady, state (NESS) of active turbulence, for the case of contractile swimmers, if $\zeta$ is sufficiently large and negative and (b) this turbulence arrests the phase separation. We quantify this suppression by showing how the coarsening-arrest length scale does not grow indefinitely, with time $t$, but saturates at a finite value at large times. We characterise the statistical properties of this active-scalar turbulence by employing energy spectra and fluxes and the spectrum of $\phi$. For sufficiently high Reynolds numbers, the energy spectrum $\mathcal E(k)$ displays an inertial range, with a power-law dependence on the wavenumber $k$. We demonstrate that, in this range, the flux $\Pi(k)$ assumes a nearly constant, negative value, which indicates that the system shows an inverse cascade of energy, even though energy injection occurs over a wide range of wavenumbers in our active-CHNS model.
\end{abstract}

\maketitle

%\newpage

\section{Introduction}
\textit{Active turbulence}, spatiotemporal chaos in active-matter systems [see, e.g., Refs.~\cite{wensink2012meso,qi2022emergence,alert2022active,mukherjee2023intermittency}], has garnered considerable attention over the past decade. This intriguing form of turbulence manifests itself in various experimental systems, including bacterial suspensions~\cite{dunkel2013fluid,kaiser2014transport,wensink2012meso,joy2020friction,linkmann2019phase,linkmann2020condensate,aranson2022bacterial,kiran2023irreversibility}, suspensions of microtubles, and molecular motors~\cite{opathalage2019self, wu2017transition}. In classical-fluid turbulence, a nonequilibrium statistically steady state (NESS) is reached when the fluid is driven by an external force; by contrast, in \textit{active} fluids, the microscopic constituents drive the system by converting chemical sources of energy into kinetic energy~\cite{dunkel2013fluid,dunkel2013minimal}. Many studies have focused on understanding emergent turbulence-type patterns by using continuum hydrodynamical models, in which phenomenological parameters depend on the microscopic details of the active fluid.
An overview of the various models can be found in Ref.~\cite{alert2022active}.
 % \sout{Numerical investigations have focused on mathematical models that elucidate active turbulence, notably including the Toner-Tu model~\cite{bratanov2015new, james2021emergence, gibbon2023analytical}, the Toner-Tu-Swift-Hohenberg (TTSH) model~\cite{kiran2023irreversibility,singh2022lagrangian}, and models addressing turbulence within active nematics, polar and spinner fluids~\cite{koch2021role, worlitzer2021motility, alert2020universal,rorai2022coexistence, chatterjee2021inertia,reeves2021emergence}}.
 %In \textit{active} fluids the microscopic constituents drive the system by converting chemical sources of energy into kinetic energy \cite{dunkel2013fluid,dunkel2013minimal} as opposed to classical fluids, where a non equilibrium steady state (NESS) is reached by driving the fluid by a external force. 
% \sout{In contrast to classical turbulence, 
% which relies on external forces to drive fluid motion, active turbulence operates uniquely by leveraging an intrinsic internal mechanism at small scales. In turn, this gives rise to the formation of coherent structures on a smaller scale. The models employed to investigate active turbulence inherently exist in a state of nonequilibrium due to the dual nature of these active entities: while they inject energy into the system at small scales, they also simultaneously dissipate energy.} 
In certain models, the energy spectrum of such turbulence exhibits universal power-law behaviors~\cite{alert2020universal,mukherjee2023intermittency}. In contrast, there are instances in which power-law exponents for these energy spectra depend on parameters in the model~\cite{bratanov2015new,joy2020friction,kiran2023irreversibility}. The elucidation of the statistical properties of these types of emergent turbulent states continues to be an important challenge in active-matter research. 
Recent studies have shown the importance of fluid inertia in some systems that display active turbulence such as active \textit{polar} and \textit{nematic} fluids [see, e.g., Refs.~\cite{thampi2013velocity,thampi2014vorticity,chatterjee2021inertia,rorai2022coexistence}]. In addition, considerable attention has been directed towards the study of \textit{scalar active fluids}, in which the intricate spatiotemporal evolution of an active fluid arises from the interaction of a scalar order parameter $\phi$ with the fluid velocity $\bm u$. Scalar active fluids are simpler than their polar or nematic counterparts, yet they are rich enough to yield intriguing emergent NESSs~\cite{saha2020scalar, tiribocchi2015active, frohoff2023non}; and they have been used in studying active droplets~\cite{padhan2023activity} and active stratified turbulence~\cite{bhattacharjee2022activity}. 

We uncover a new type of turbulence -- activity-induced homogeneous and isotropic turbulence in a model that has been employed to investigate motility-induced phase separation (MIPS)~\cite{tiribocchi2015active, padhan2023activity} in a system of microswimmers. The active Cahn-Hilliard-Navier-Stokes equations (CHNS), also known as the active model H ~\cite{tiribocchi2015active}, provides a natural theoretical framework for our study. In this CHNS model, a single scalar order parameter $\phi$ [which is positive (negative) in regions where the microswimmer density is high (low)] is coupled with the velocity field $\bm u$. The activity of the microswimmers is governed by an activity parameter $\zeta$ that is positive for \textit{extensile} swimmers and negative for \textit{contractile} swimmers.  With extensile swimmers, this system undergoes complete phase separation, which is similar to that in binary-fluid mixtures~\cite{tiribocchi2015active}. By carrying out pseudospectral direct numerical simulations (DNSs) we show, for the first time, that this model develops an emergent nonequilibrium, but statistically steady, state (NESS) of active turbulence, for the case of contractile swimmers, if $\zeta$ is sufficiently large and negative. This turbulence arrests the phase separation into regions with positive and negative values of $\phi$, in much the same way as conventional fluid turbulence leads to the suppression of phase separation in a binary-fluid mixture~\cite{perlekar2014spinodal,perlekar2017two,perlekar2019kinetic}.
We quantify this suppression by showing how the coarsening-arrest length scale does not grow indefinitely, with time $t$, but saturates at a finite value at large times. We then characterise the statistical properties of this active-scalar turbulence by employing the energy spectrum and fluxes, which are familiar from classical fluid turbulence, and also the spectrum of $\phi$, which is used in studies of phase separation. For sufficiently high Reynolds numbers, the energy spectrum $\mathcal{E}(k)$ displays an inertial range, with a power-law dependence on the wavenumber $k$. We demonstrate that, in this range, the flux $\Pi(k)$ assumes a nearly constant, negative value, which indicates that the system shows an inverse cascade of energy that is similar to its counterpart in 2D homogeneous and isotropic fluid turbulence, even though energy injection occurs over a wide range of wavenumbers in our active-CHNS model. 
%Our results are of potential relevance to systems of contractile swimmers such as \textit{Chlamydomonas reinhardtii}~\cite{yeomans2014introduction,fragkopoulos2021self} (\textit{C. reinhardtii}) and synthetic active colloids ~\cite{zottl2016emergent,howse2007self}.

%We demonstrate that in the limit of low activity, the energy spectra exhibit non-universal scaling behaviour. However, beyond a critical activity, inertia becomes prominent and drives towards a turbulence dominated by inertial effects leading to universal scaling exponents in the energy spectrum $\mathcal E(k) \sim k^{-\delta}$ with $\delta \simeq 5/3$; the exponent is similar to that of the classical two-dimensional (2D) turbulence where it shows Kolmogorov scaling exponent $5/3$. Furthermore, our analysis, which includes scale-by-scale energy contributions and energy flux calculations, demonstrates that energy injection occurs across all length scales due to the active stress, and the fluid inertia plays a pivotal role in transferring this energy across scales. Together, these factors are key drivers of the observed turbulence in our study.\\

The remaining part of this paper is organised as follows. Section~\ref{sec:MMSM} introduces the active CHNS model, summarises the numerical methods we employ to study it, and defines the statistical measures we use to characterise active turbulence in this model. In Section~\ref{sec:results} we present the results of our study. We discuss the significance of our results in Section~\ref{sec:conclusions}.
%%%%%%%%%%%%%%%%%%%%%%%%%%%%%%%%%%%
\begin{figure}[!]
{   
    \includegraphics[width=0.49\textwidth]{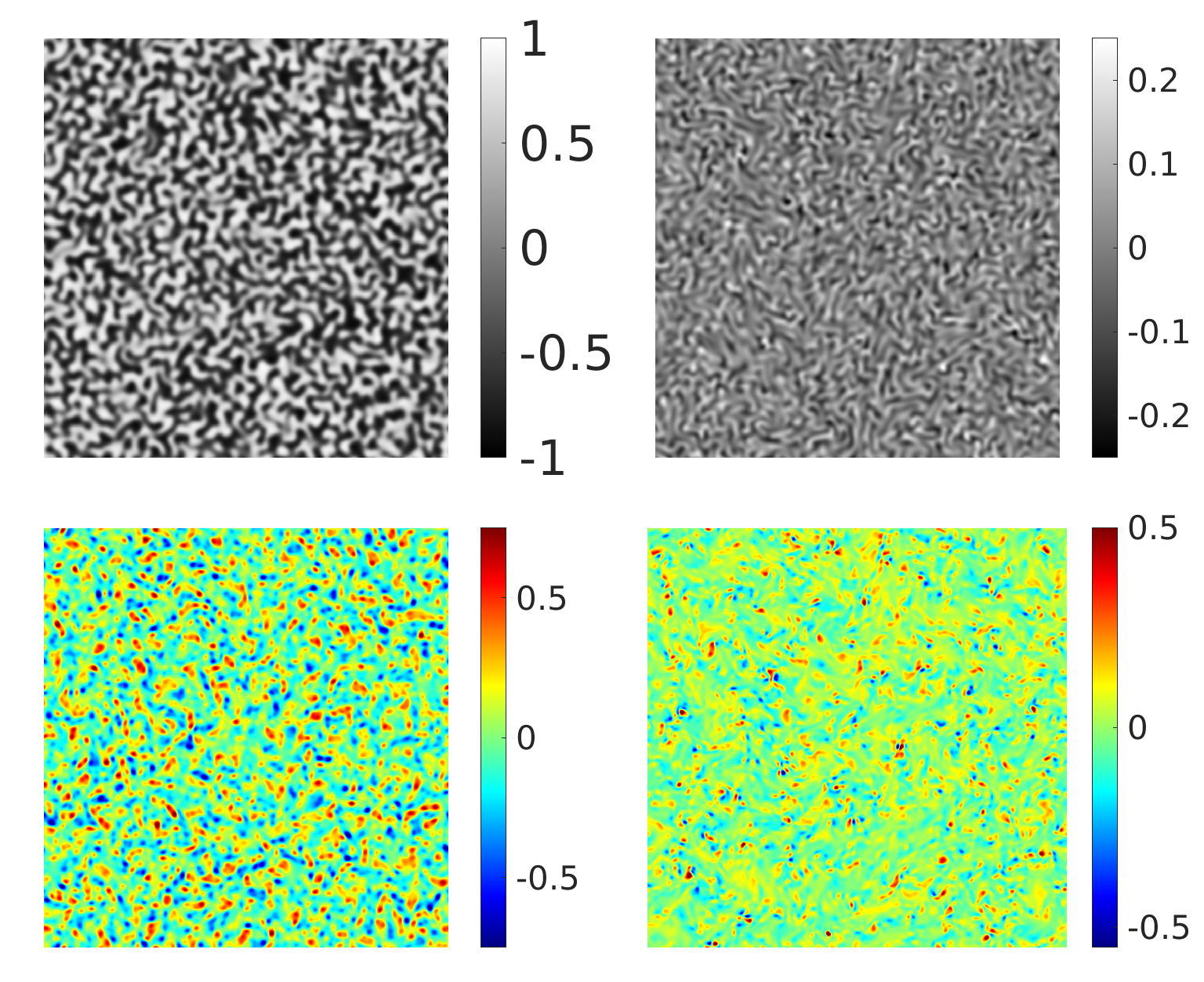}
    \put(-250, 100){\rm {\bf(c)}}
    \put(-120, 100){\rm {\bf(d)}}
    \put(-250, 205){\rm {\bf(a)}}
    \put(-120, 205){\rm {\bf(b)}}
}     
    \caption{Pseudo-gray-scale plots of the $\phi$ field [at representative times in the nonequilibrium statistically steady state (NESS)] for the activity parameter (a) $|\zeta| = 0.01$ and (b) $|\zeta| = 1.5$. Pseudocolor plots of the vorticity field, normalized by the maximum of $|\omega|$, are shown in (c) and (d) for the parameters in (a) and (b), respectively.}
    \label{fig:pcolor}
\end{figure}
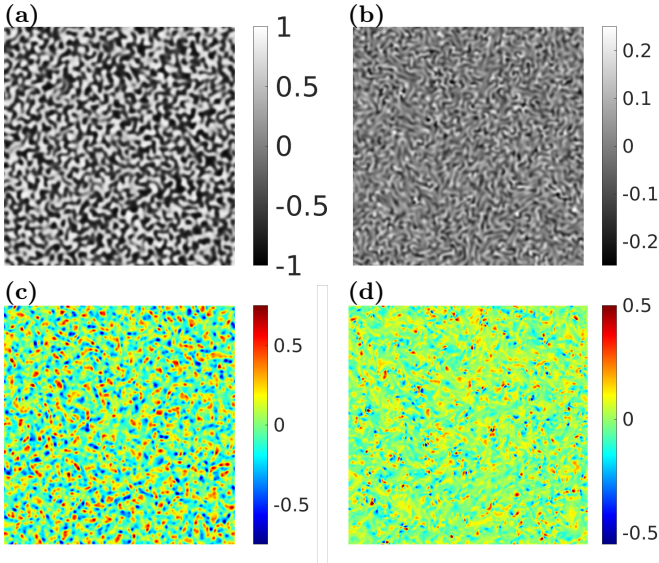
%%%%%%%%%%%%%%%%%%%%%%%%
%%%%%%%%%%%%%%%%%%%%%%%%%%%%%%%%%%%
\begin{figure*}[!]
{   
    %\hspace{-0.5cm}
    % \includegraphics[width=0.425\textwidth]{images/vort_pcolor.png}
    \includegraphics[width=0.475\textwidth]{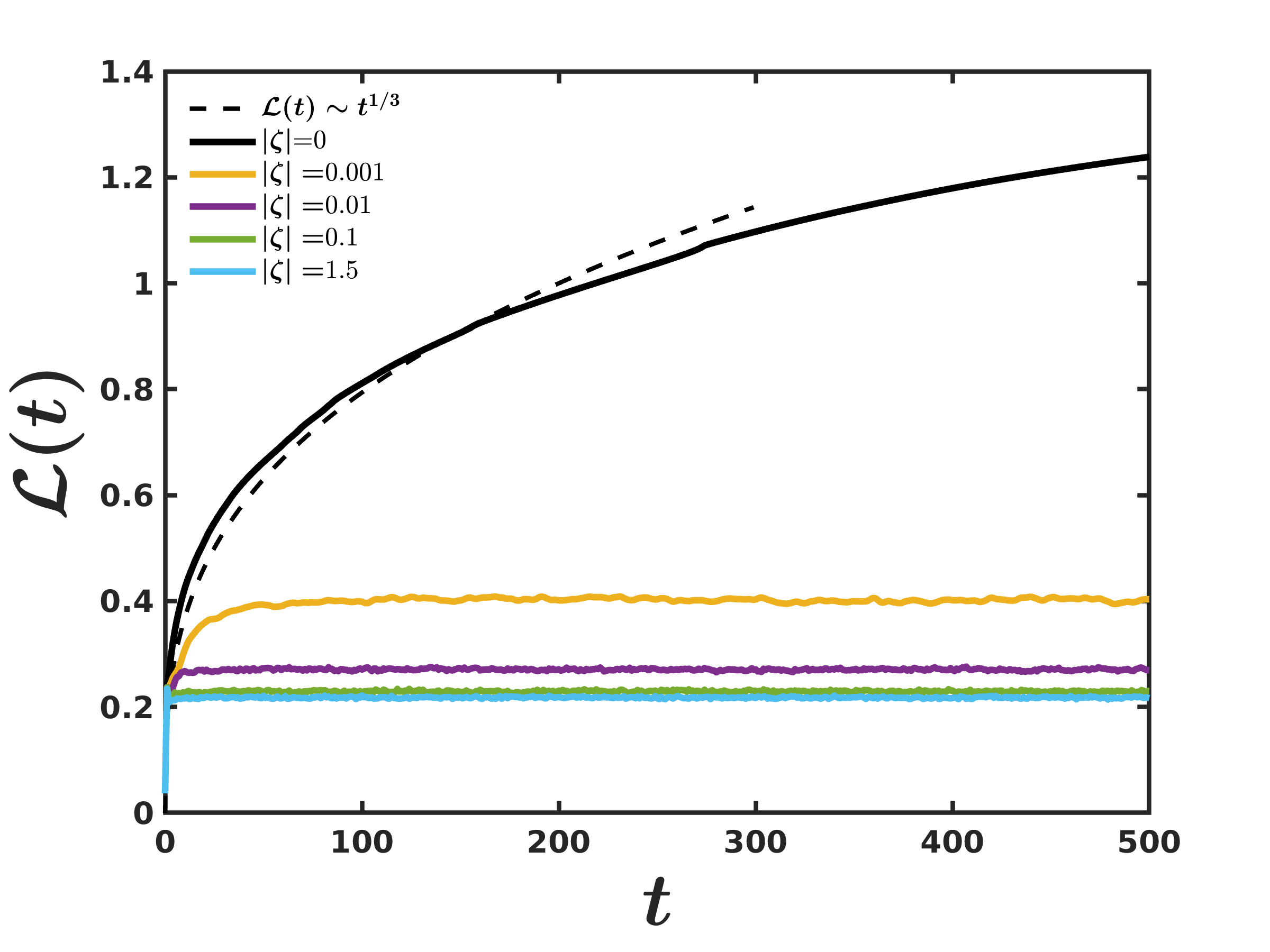}
    \includegraphics[width=0.475\textwidth]{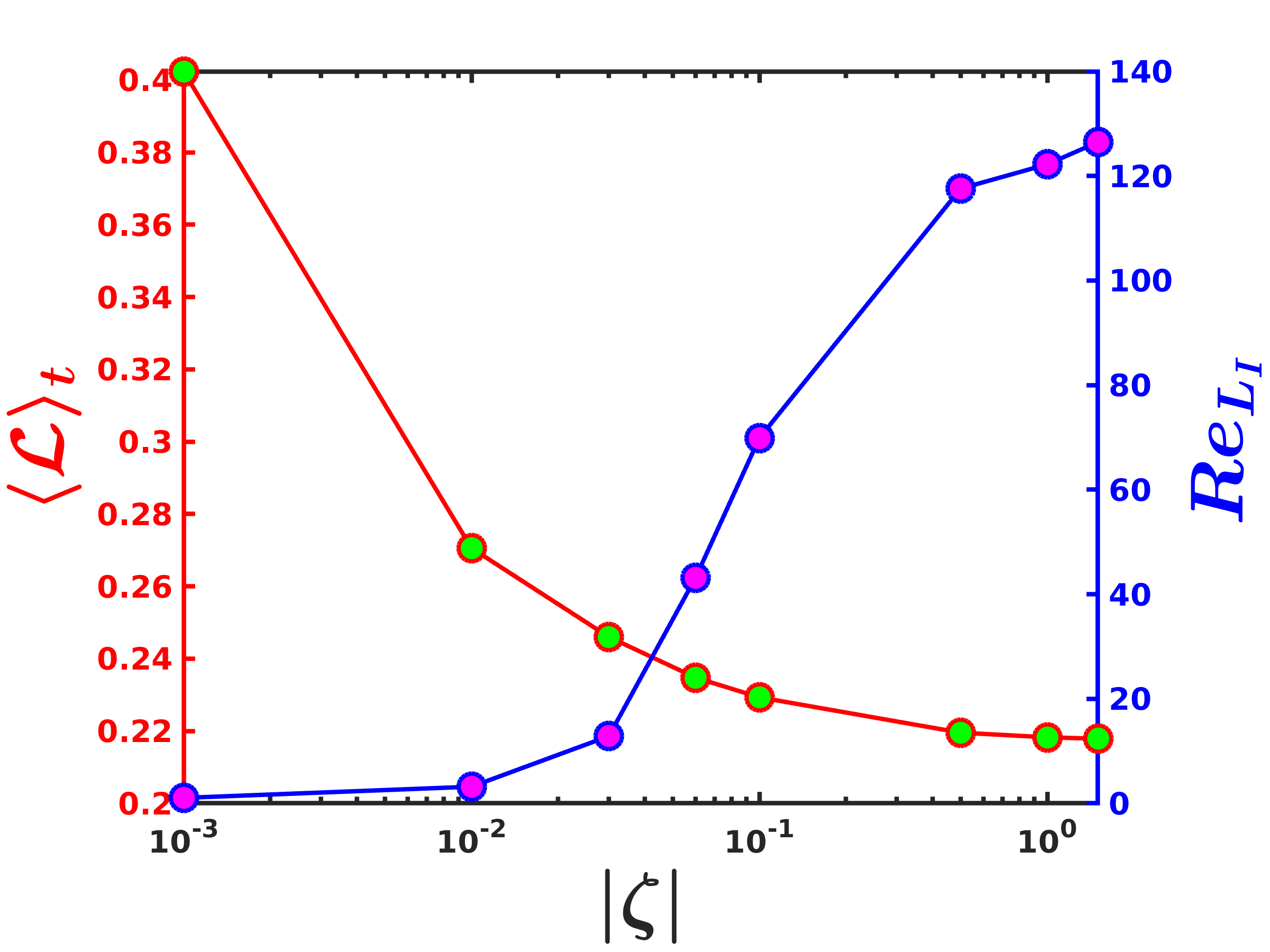}
    \put(-500, 170){\rm {\bf(a)}}
    \put(-250, 170){\rm {\bf(b)}}
    %\put(-160, 100){\rm {\bf(c)}}
}     
    \caption{(a) Plot of the coarsening length scale $\mathcal L(t)$  [Eq.~(\ref{eq:urmsLs})] versus time $t$ for various values of $|\zeta|$; the plot for $\zeta = 0$ shows growth that is consistent with the Lifshitz-Slyozov form $\mathcal L(t) \sim t^{1/3}$ (dashed line); $\mathcal L(t)$ saturates to a finite value for $|\zeta| > 0$. (b) Log-linear plots of the mean coarsening-arrest scale $L_c = \langle\mathcal L(t)\rangle_t$ (red curve) and the integral-scale Reynolds number $Re_{L_I}$ (blue curve) versus $|\zeta|$.}
    \label{fig:domain_growth}
\end{figure*}
%%%%%%%%%%%%%%%%%%%%%%%%

\section{Model, Methods, and Statistical Measures}
\label{sec:MMSM}

We introduce the active-CHNS model in Subsection~\ref{subsec:model}. In Subsection~\ref{subsec:statistics} we describe the statistical measures we use to characterise active turbulence in this model. Finally, in Subsection~\ref{subsec:DNS} we give the details of our pseudospectral DNS.

\subsection{The Active Cahn-Hilliard-Navier-Stokes Model}
\label{subsec:model}

We consider the incompressible active CHNS equations (also called active model H) to study active turbulence in systems of contractile swimmers~\cite{tiribocchi2015active,padhan2023activity} in two spatial dimensions (2D):
\begin{eqnarray}
    \partial_t \phi + (\bm u \cdot \nabla) \phi &=& \text{M} \nabla^2\left(\frac{\delta \mathcal F}{\delta \phi}\right)\, ;\label{eq:ch}\\ 
    \partial_t \omega + (\bm u \cdot \nabla) \omega &=& \nu \nabla^2\omega + \frac{3}{2}\epsilon \nabla \times (\nabla \cdot \bm \Sigma^A) - \alpha \omega\, ;\;\label{eq:ns}\\
    \nabla \cdot \bm u &=& 0\,; \label{eq:incomp} 
\end{eqnarray}
$\omega$ is the vorticity field; $\nu$, $\alpha$, and M are the kinematic viscosity, bottom friction, and mobility, respectively. 
$\mathcal F$ is the Landau-Ginzburg variational free-energy functional
\begin{eqnarray}
     \mathcal F[\phi, \nabla \phi] = \int_{\Omega} \left[\frac{3}{16} \frac{\sigma}{\epsilon}(\phi^2-1)^2 + \frac{3}{4} \sigma \epsilon |\nabla \phi|^2\right]\,,\label{eq:functional}
\end{eqnarray}
in which the first term is a double-well potential with minima at $\phi = \pm 1$. The scalar order parameter $\phi$ is positive (negative) in regions where the microswimmer density is high (low); in the interfaces between these regions, $\phi$ varies smoothly, over a width $\epsilon$. The free-energy penalty for an interface is given by the bare surface tension $\sigma$. In the inherently nonequilibrium active model H all
terms in the stress tensor do not follow from $\mathcal F$. In particular, we must include the stress tensor $\bm \Sigma^A$, which has the form of a nonlinear Burnett term and has the components~\cite{padhan2023activity,tiribocchi2015active,bhattacharjee2022activity,das2020transition}
\begin{eqnarray}
    \Sigma^{A}_{ij} = -\zeta \left[\partial_i \phi \partial_j \phi - \frac{\delta_{ij}}{2} |\nabla \phi|^2\right]\,,
    \label{eq:tensor}
\end{eqnarray}
where $\zeta$, the activity coefficient, can take both positive and negative values: $\zeta < 0$ ($\zeta > 0$) for contractile (extensile) swimmers~\cite{tiribocchi2015active}. 

\subsection{Statistical Characterisation}
\label{subsec:statistics}

To characterise the statistical properties of active-scalar turbulence, we employ energy spectra and fluxes, 
which are familiar from classical fluid turbulence, and the spectrum of $\phi$. These quantities not only help us to understand, via DNS, the emergent turbulent like behaviour (characterized by spatiotemporal fluctuations) in Eqs. (1-3), but they also aid us in differentiating active scalar turbulence from classical 2D incompressible fluid turbulence [see, e.g., Refs.~\cite{boffetta2012two,pandit2017overview,kraichnan1967inertial,kraichnan1980two}] and also turbulent patterns found in other active systems [see, e.g., Refs.~\cite{thampi2013velocity,thampi2014vorticity,chatterjee2021inertia,rorai2022coexistence}]. We define the shell-averaged energy and phase-field spectra, respectively,  both at time $t$ and averaged over time [the time average is denoted by $\langle \cdot \rangle_t$]:
\begin{eqnarray}     
     \mathcal{E}(k,t) &\equiv& \sum_{k\le k' <k+1} |\hat{\bm u}(\bm k', t)|^2\,; ~~~\Phi(k,t) \equiv \sum_{k\le k' <k+1} |\hat{\phi}(\bm k', t)|^2\,; \nonumber \\
      \mathcal{E}(k) &\equiv& \langle \mathcal{E}(k,t) \rangle_t \,; ~~~ \Phi(k) \equiv \langle \Phi(k,t) \rangle_t \,;
      \label{eq:spectra}
\end{eqnarray}
%%%%%%%%%%%%%%%%%%%%%%%%%%%%%%%%%%%
\begin{figure*}[!]
{
   \includegraphics[width=0.475\textwidth]{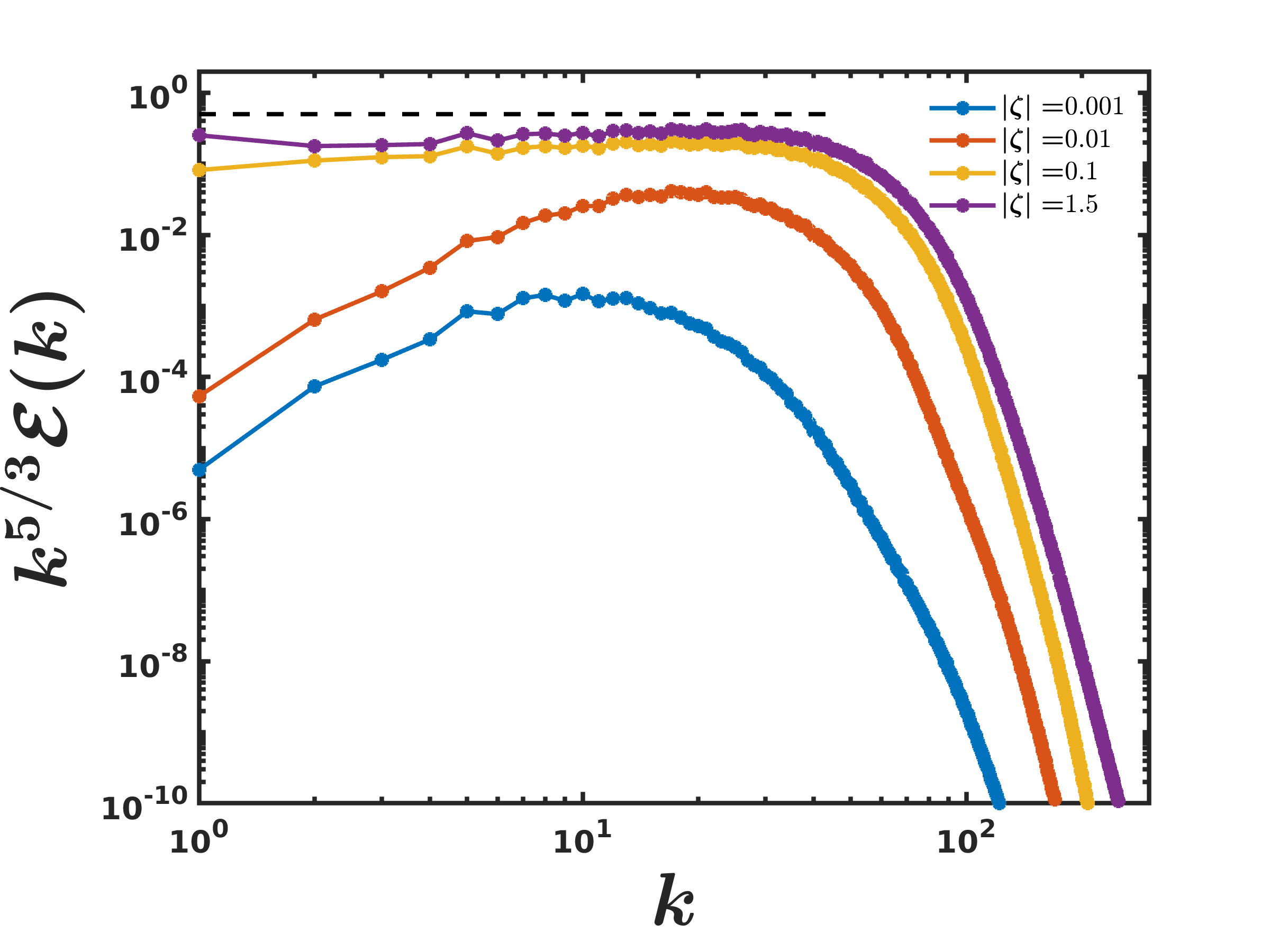}
   \includegraphics[width=0.475\textwidth]{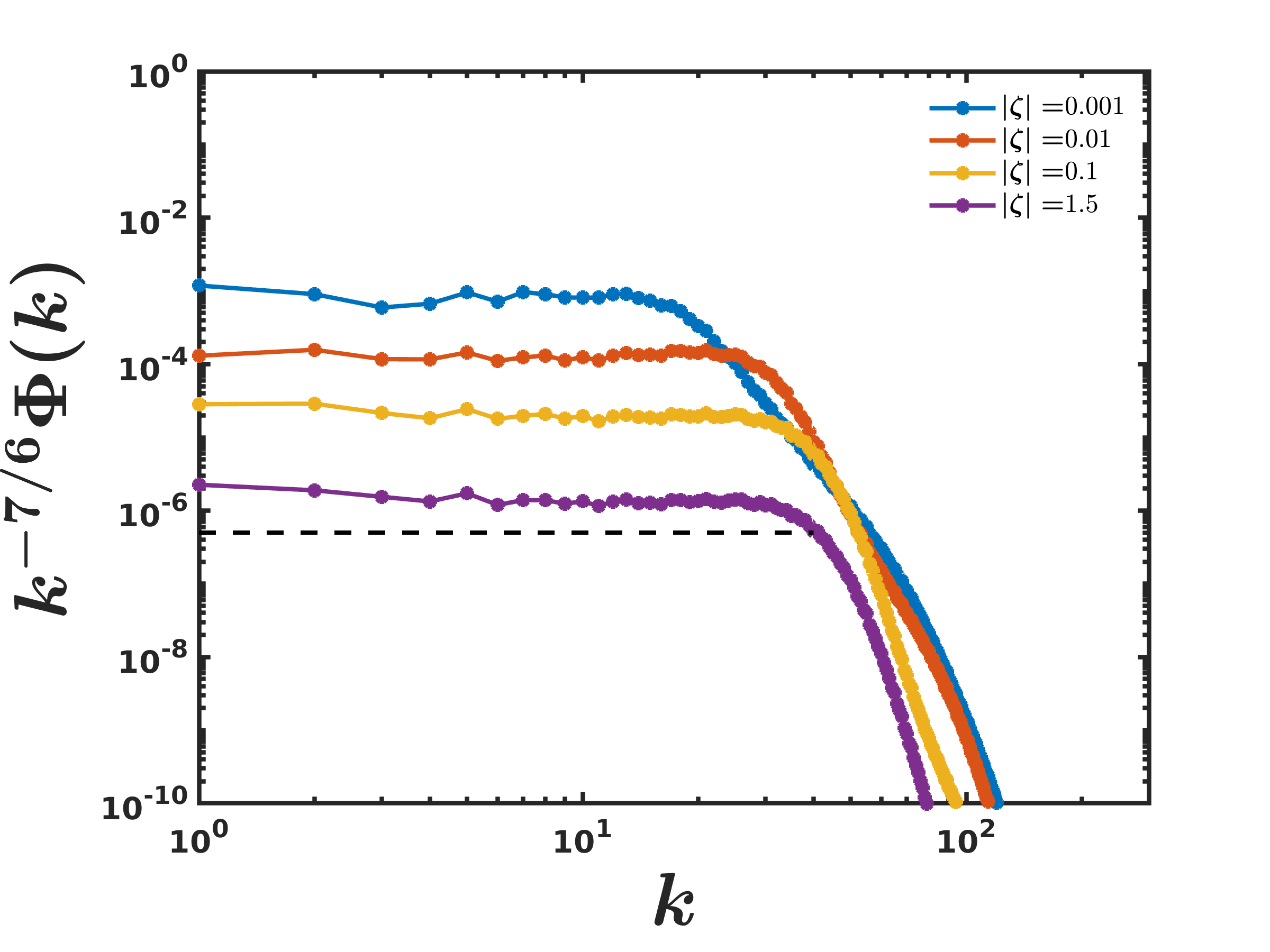}
   \put(-500, 170){\rm {\bf(a)}}
   \put(-250,170){\rm {\bf(b)}}
}     
    \caption{Log-log plots versus the wavenumber $k$ of spectra for $|\zeta| = 0.001, 0.01, 0.1, 1.5$: (a) The compensated energy spectrum $k^{5/3}\mathcal{E}(k)$. For $|\zeta|=0.1, 1.5$, the exponent is $-5/3$ (black dotted line in the compensated spectrum). (b) The compensated phase-field spectrum $k^{-7/6}\Phi(k)$ for $|\zeta| = 0.001, 0.01, 0.1, 1.5$. These spectra show the power-law behaviour $\Phi(k) \sim k^{7/6}$ (black dotted line in the compensated spectrum).}
    \label{fig:spectra}
\end{figure*}
%%%%%%%%%%%%%%%%%%%%%%%%%%%%%%%%%%%
the caret denotes a spatial Fourier transform. Our CHNS study of active scalar turbulence uses the Reynolds, P\'eclet, Weber, and the non-dimensional friction numbers that are, respectively: 
\begin{eqnarray}
\text{Re} &=& \frac{\text{L}_I u_{rms}}{\nu}\,;\;\; Pe = \frac{\epsilon {L}_I u_{rms}}{M \sigma}\,;\nonumber \\
We &=& \frac{L_I u_{rms}^2}{\sigma}\,;\;\; \alpha' = \frac{\alpha L_I}{u_{rms}}\,;
\label{eq:nondimnos}
\end{eqnarray}
$u_{rms}$ and $L_I$ are, respectively, the root-mean-squared velocity and the fluid integral length scale; and $\mathcal L(t)$ 
and $L_c$ are the time-dependent and mean coarsening length scales; these are defined as follows:
\begin{eqnarray}
    u_{rms} &=& \left[\sum_k \mathcal{E}(k)\right]^{1/2}\;;\;\;\; L_I = 2\pi \frac{\sum_k\mathcal E(k)}{\sum_k k \mathcal E(k)}\;;\nonumber \\
    \mathcal L(t) &=& \frac{\sum_k \Phi(k,t)}{\sum_k k \Phi(k,t)}\;;\;\;\; L_c = \langle \mathcal L(t) \rangle_t\;.
     \label{eq:urmsLs}
\end{eqnarray}
In Table~\ref{tab:param}, we provide the non-dimensional parameters from our direct numerical simulations (DNSs) for various values of the activity parameter $|\zeta|$. Furthermore, $\mathcal{E}(k,t)$ satisfies the following energy-budget equation~\cite{perlekar2019kinetic,alexakis2018cascades,verma2019energy}:
\begin{equation}
    \partial_t \mathcal E(k, t) = T(k, t) + D_{\alpha}(k, t) + D_{\nu}(k, t) + S^{\phi}(k, t)\,, \label{eq:budget}
\end{equation}
where
\begin{eqnarray}
    T(k, t) &=& - \Re\left[\sum_{k\leq |\bm k'|< k+1} [ \hat{\bm u}(-\bm k', t)\cdot \bm P(\bm k') \cdot \widehat{(\bm u \cdot \nabla \bm u)} (\bm k', t)]\right], \nonumber \\
    D_{\alpha}(k, t) &=& -2 \alpha \mathcal E(k,t),~~~~D_{\nu}(k, t) = -2 \nu k^2 \mathcal E(k,t), \nonumber \\
    S^{\phi}(k, t) &=& \Re\left[ \sum_{k\leq |\bm k'|< k+1} [ \hat{\bm u}(-\bm k', t)\cdot \bm P(\bm k') \cdot \widehat{(\nabla \cdot \bm \Sigma^A)} (\bm k', t)]\right]
    \label{eq:contributions}
\end{eqnarray}
are, respectively, the energy transfer because of the inertial term, the energy dissipations arising from friction and the viscosity, and the energy transfer via the active-stress term; the transverse projector $\bm P(\bm k)$, which  enforces the incompressibility condition, has the components $P_{ij} \equiv (\delta_{ij} - {k_ik_j}/{k^2})$ . We also use the following mean energy transfers from the inertial, friction, viscous, and active-stress terms, and the associated kinetic-energy and active-stress fluxes~\cite{verma2019energy}:
\begin{eqnarray}
T(k) &=& \langle T(k, t)\rangle_t\,;\;\;\;\; S^{\phi}(k) = \langle S^{\phi}(k, t)\rangle_t\,;\;\;\; \nonumber \\
D_{\alpha}(k) &=& \langle D_{\alpha}(k, t)\rangle_t\,;\;\;\;\; D_{\nu}(k) = \langle D_{\nu}(k, t)\rangle_t\,;\nonumber \\
  \Pi(k) &=& -\int_{0}^{k'} T(k') dk'\,;\;\;\;\; \Pi^{\phi}(k) = -\int_{0}^{k'} S^{\phi}(k') dk'\,.
  \label{eq:fluxes}
\end{eqnarray}

\subsection{Direct Numerical Simulation}
\label{subsec:DNS}

We carry out DNSs of the active-CHNS partial differential equations (\ref{eq:ch})-(\ref{eq:tensor}) by using the pseudospectral method~\cite{canuto2012spectral,perlekar2017two,Pal_2016,padhan2023activity} in a 2D periodic square domain, $\mathcal{D}\equiv[0,L]^{2}$, with $L$ the length of the side of the square. We evaluate spatial derivatives in Fourier space and the nonlinear terms in physical space. For time integration, we use the semi-implicit exponential-time-difference Runge-Kutta2 (ETDRK2) method~\cite{cox2002exponential}. We employ the $1/2$-dealiasing scheme to remove the Fourier aliasing errors~\cite{canuto2012spectral,perlekar2017two,Pal_2016,padhan2023activity}. To resolve the interface of width $\epsilon$, we ensure that there are three grid points across the interface. We use a CUDA-C code that we have developed and optimised for an
NVIDIA A100 processor.\\
%\section{Initial conditions}

\subsection{Initial conditions}
\label{subsec:IC}

We use the following initial conditions for the $\omega$ and $\phi$ fields:
\begin{equation}
    \omega(x, y, 0) = 0 \, ; ~~~~    \phi(x, y, 0) = \phi_0 + \xi(x,y)\,;
\end{equation}
$\xi(x,y)$, a random number distributed uniformly on the interval  $[-0.1, 0.1]$, provides a random perturbation to the $\phi = \phi_0 = 0$ state.

% \section{Non-dimensional parameters}
% For our study on active scalar turbulence, we use the dimensionless Reynolds, Peclet, Weber, and the non-dimensional friction numbers that are, respectively: 
% \begin{eqnarray}
% \text{Re} &=& \frac{\text{L}_I u_{rms}}{\nu}\,;\;\; Pe = \frac{\epsilon {L}_I u_{rms}}{M \sigma}\,;\nonumber \\
% We &=& \frac{L_I u_{rms}^2}{\sigma}\,;\;\; \alpha' = \frac{\alpha L_I}{u_{rms}}\,;
% \label{eq:nondimnos}
% \end{eqnarray}
% $u_{rms}$ and $L_I$ are the root-mean-sqare velocity and the inegral length scale defined as 
% \begin{eqnarray}
%     u_{rms} &=& \displaystyle\sqrt{\displaystyle\sum_k \mathcal{E}(k)} \nonumber\\
%     L_I &=& 2\pi {\displaystyle\sum_k\mathcal E(k)}/{\displaystyle \sum_k k \mathcal E(k)}.
% \end{eqnarray}
% We provide the non-dimensional parameters and other essential parameters used in our simulations in Table~\ref{tab:param}.

\section{Results}
\label{sec:results}

In this Section we present a series of DNSs that we have designed to demonstrate how the activity of contractile swimmers [$\zeta < 0$ in Eq.~(\ref{eq:tensor})] leads to active turbulence that is strong enough to suppress motility-induced phase separation. Our results are the active-turbulence counterparts of the suppression of phase separation (also called coarsening arrest) by fluid turbulence [see, e.g., Refs.~\cite{perlekar2014spinodal,perlekar2017two}].

%%%%%%%%%%%%%%%%%%%%%%%%%%%%%%%%%%%
\begin{figure*}[!]
{    \includegraphics[width=0.475\textwidth]{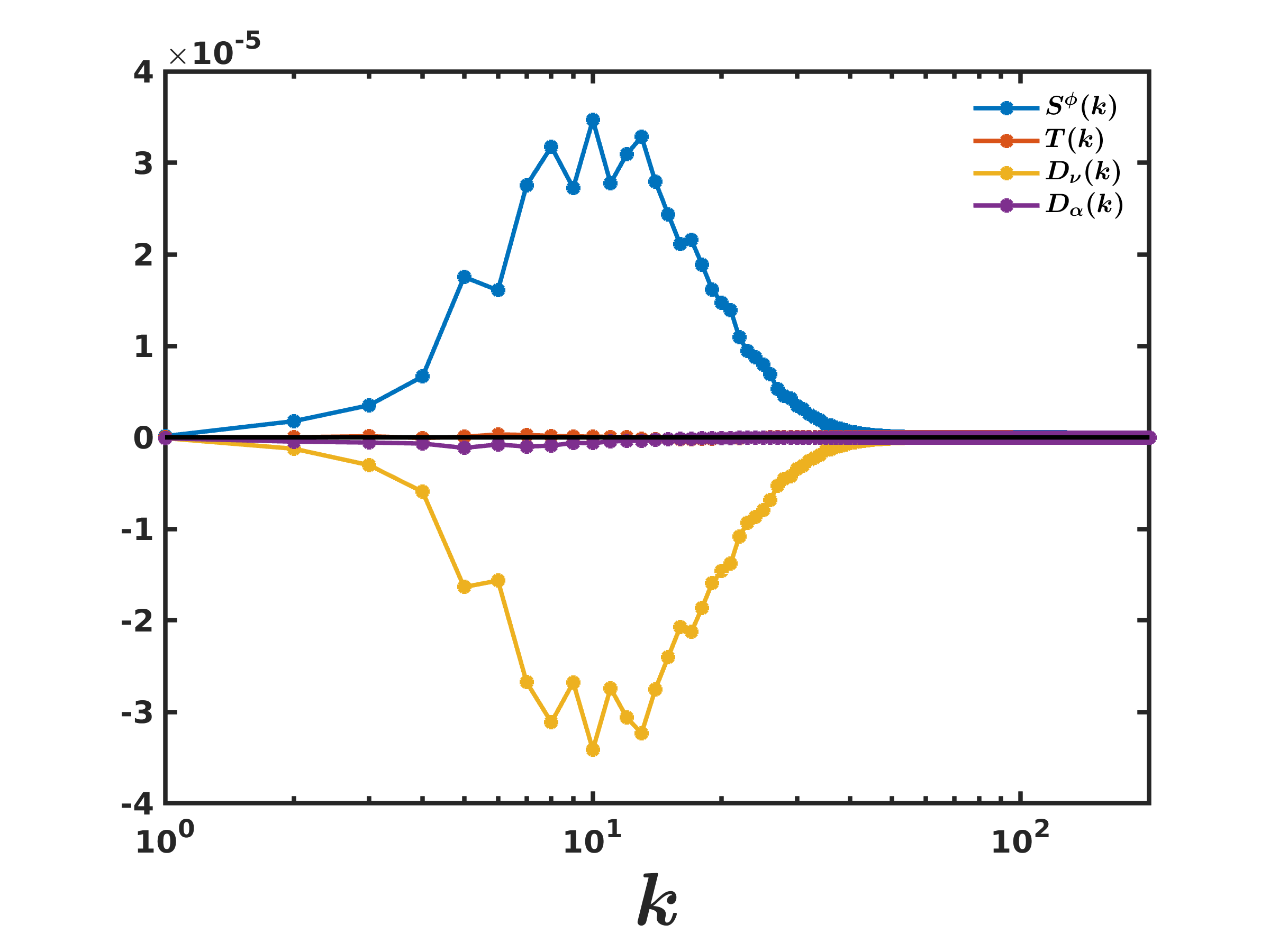}
   \includegraphics[width=0.475\textwidth]{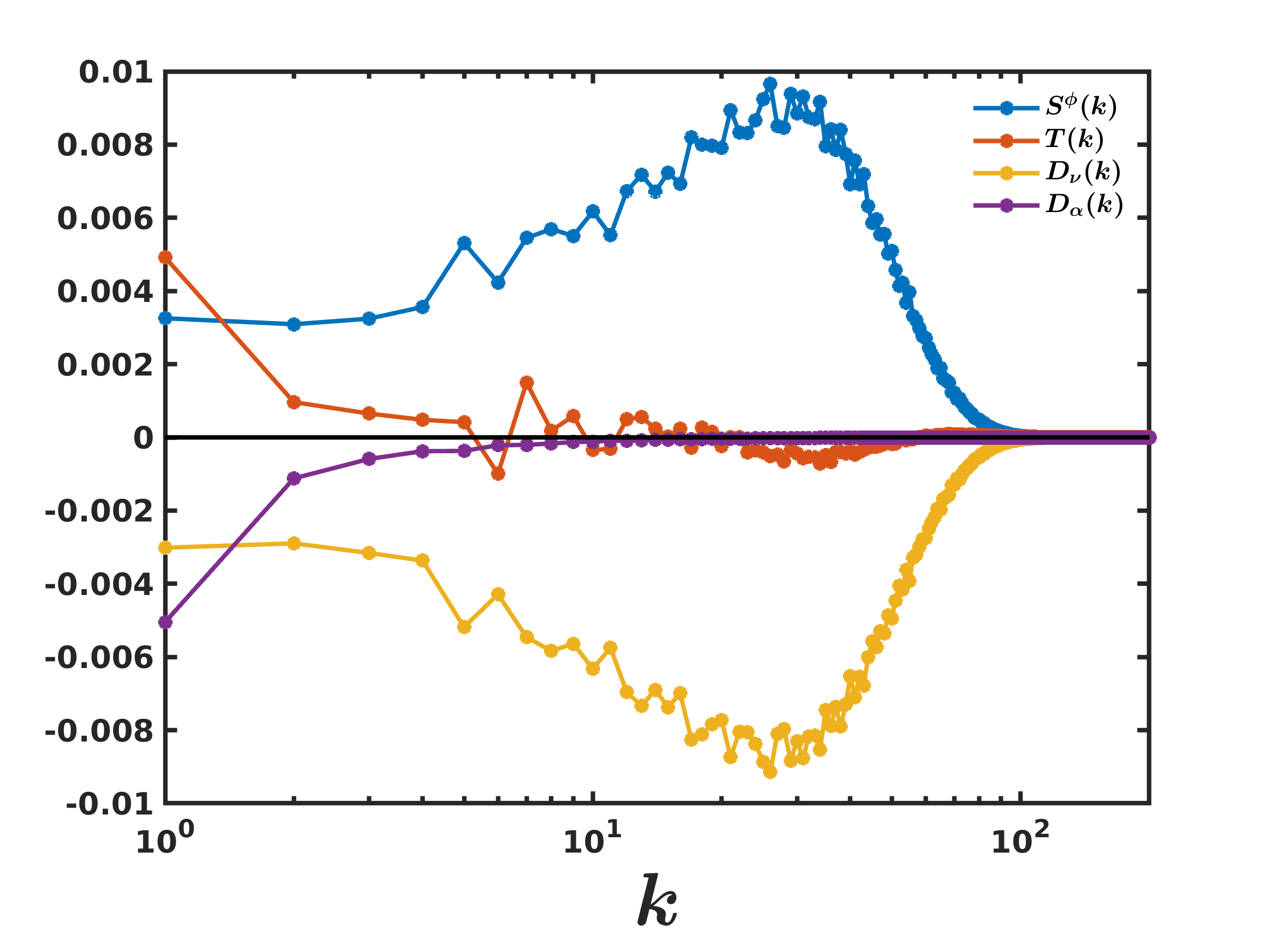}
    \put(-500, 170){\rm {\bf(a)}}
    \put(-250, 170){\rm {\bf(b)}}
}     
    \caption{Plots versus $k$ (log scale) of the contributions $ T(k),\; D_{\alpha}(k),\; D_{\nu}(k),\; \rm{and}\; S^{\phi}(k)$, in red, purple, yellow, and blue, respectively, for (a) $|\zeta| = 0.001$ and (b) $|\zeta| = 1.5$.}
    \label{fig:budget}
\end{figure*}
%%%%%%%%%%%%%%%%%%%%%%%%%%%%%%%%%%%
%%%%%%%%%%%%%%%%%%%%%%%%%%%%%%%%%%%
\begin{figure*}[!]
{
    \includegraphics[width=0.32\textwidth]{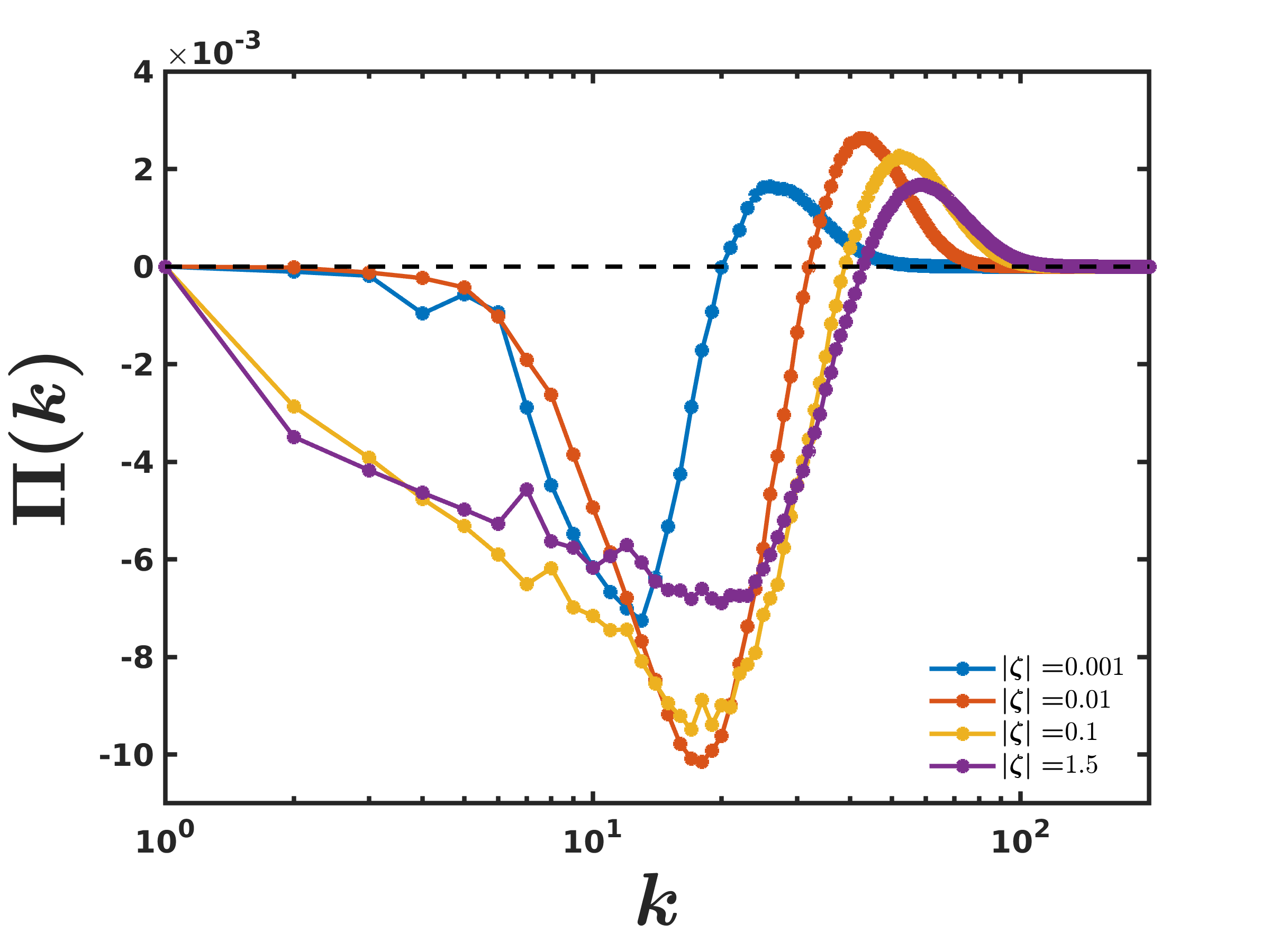}
    \includegraphics[width=0.32\textwidth]{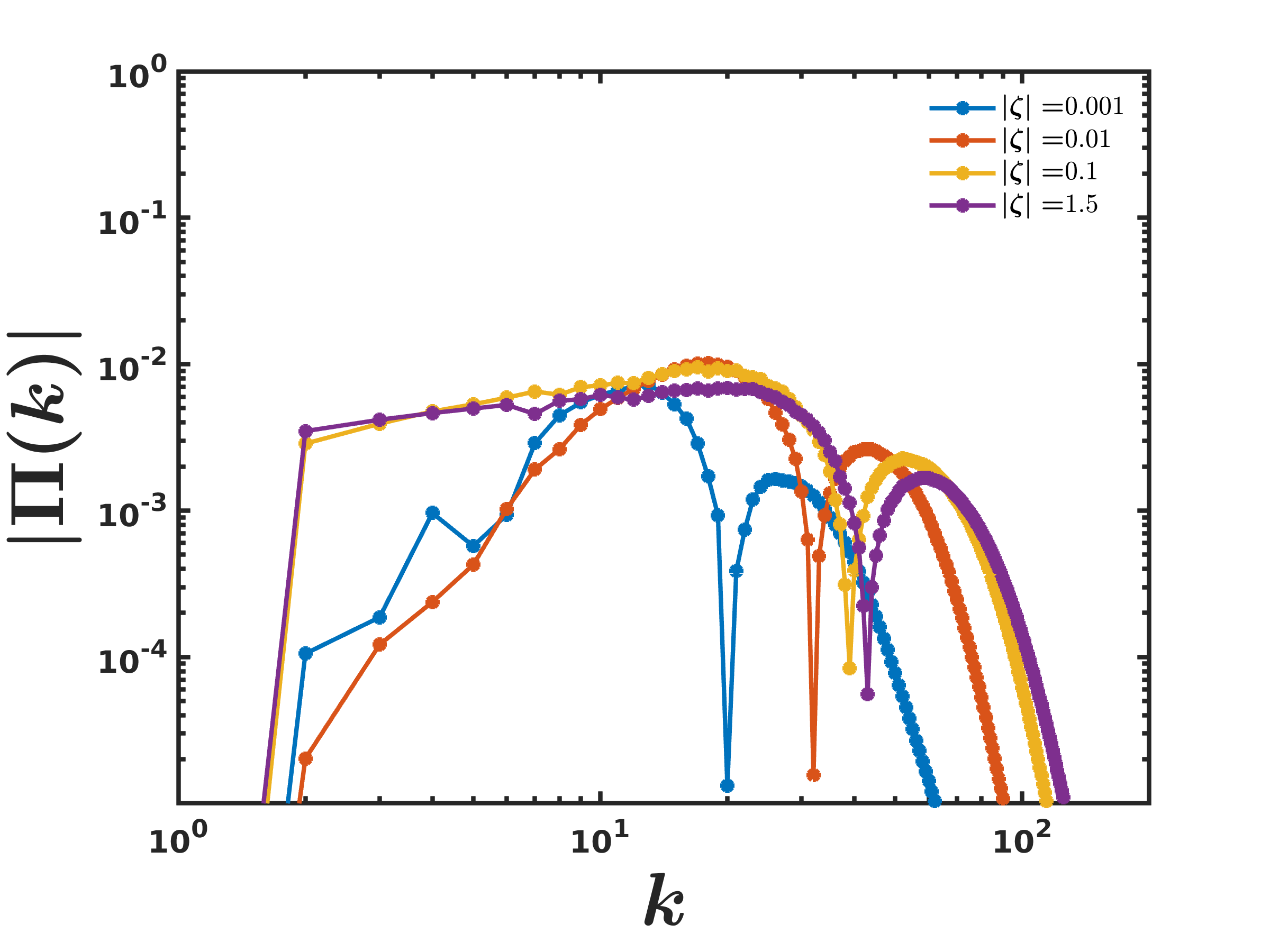}
    \includegraphics[width=0.32\textwidth]{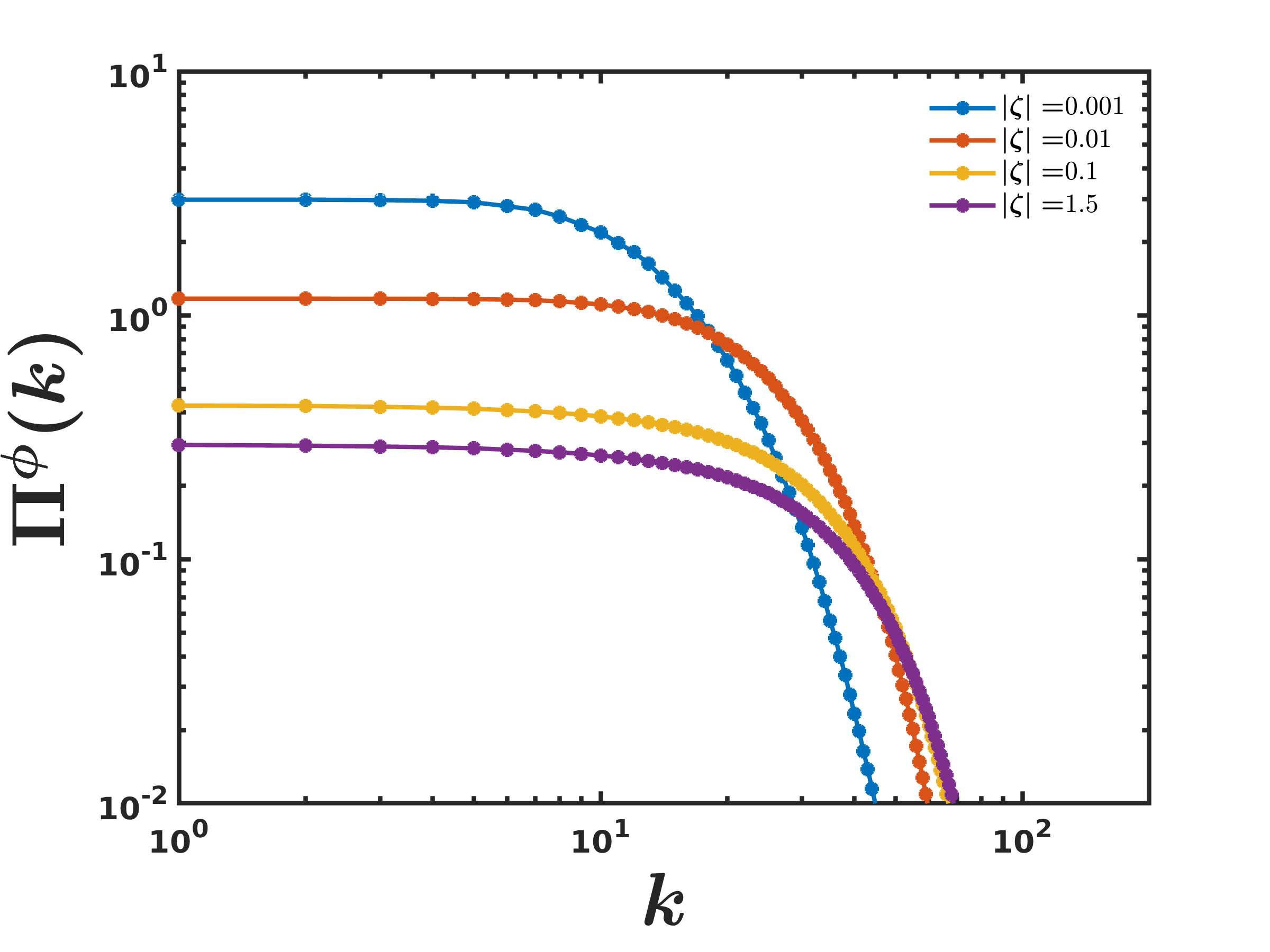}
    \put(-510, 110){\rm {\bf(a)}}
    \put(-340, 110){\rm {\bf(b)}}
    \put(-170, 110){\rm {\bf(c)}}
}     
    \caption{ Plots for $|\zeta| = 0.001, 0.01, 0.1,$ and $1.5$  of the normalized (a) energy flux $\Pi(k)L_{I}/u^{3}_{rms}$  versus $k$
    (log scale) and log-log plots versus $k$ of (b) $|\Pi(k)|L_{I}/u^{3}_{rms}$ and (c) $\Pi^{\phi}(k)L_{I}/u^{3}_{rms}$. }
    \label{fig:flux}
\end{figure*}
%%%%%%%%%%%%%%%%%%%%%%%%%%%%%%%%%%%
%%%%%%%%%%%
% \begin{figure*}
% {
%     \includegraphics[width=0.475\textwidth]{images/arrested_length_scale.png}
%     \includegraphics[width=0.475\textwidth]{images/Re_zeta.png}
%     \put(-460, 150){\rm {\bf(a)}}
%     \put(-230, 150){\rm {\bf(b)}}
% }     
%     \caption{(a)The plot of the length scale of domain-coarsening $\mathcal L(t)$ as a function of time illustrated for different values of $|\zeta|$. For $\zeta = 0$, the growth follows the power-law $\mathcal L(t) \sim t^{1/3}$. (b) The coarsening arrested length scale $L_c \equiv \langle\mathcal L(t)\rangle_t$ as a function of $|\zeta|$. The time average is taken in the statistically steady state.}
%     \label{fig:domain_growth}
% \end{figure*}

We note that, if the activity $\zeta = 0$, the stress tensor~(\ref{eq:tensor}) vanishes, so the coupled Eqs.~(\ref{eq:ch})-(\ref{eq:tensor}) decouple into Cahn-Hilliard equations or model B~\cite{hohenberg1977theory,puri2009kinetics} and the Navier Stokes equations~\cite{navier1823memoire,stokes1901mathematical,doering1995applied} for a Newtonian fluid; the advection term for the $\phi$ field is set to zero by virtue of the initial conditions. Therefore, the domain growth or coarsening takes place solely via diffusion, without hydrodynamical
effects, and it follows the well-established Lifshitz-Slyozov domain-growth form $\mathcal L(t) \sim t^{1/3}$ [see, e.g., Refs.~\cite{lifshitz1961kinetics,puri2009kinetics,bray2002theory}]; complete phase separation also occurs for extensile swimmers~\cite{tiribocchi2015active} that lead to $\zeta > 0$.

%We characterize such growth by defining the following integral length scale:
%\begin{eqnarray}
%    \mathcal L(t) = \frac{\displaystyle \sum_k \Phi(k)}{\displaystyle \sum_k k \Phi(k)}\,;
%\end{eqnarray}
%where, $\Phi(k) = \displaystyle \sum_{k \le |k'| < k+1} |\hat{\phi}(k', t)|^2$ is the phase-field spectrum and $k$ is the wave %number. 
 
 We concentrate on active-turbulence-induced suppression of phase separation and the diffusive Lifshitz-Slyozov coarsening in our model~(\ref{eq:ch})-(\ref{eq:tensor}) with contractile swimmers, for which the activity parameter $\zeta < 0$ \footnote{At \textit{low inertia} motility-induced phase separation is also suppressed in this model~\cite{tiribocchi2015active}.}.
Active turbulence and coarsening arrest in the active-CHNS model~(\ref{eq:ch})-(\ref{eq:tensor}) can be visualized qualitatively by using pseudo-gray-scale plots of the $\phi$ field as we show in Figs.~\ref{fig:pcolor} (a) and (b) for $|\zeta| = 0.01$ and (b) $|\zeta| = 1.5$, respectively, at representative times in the nonequilibrium statistically steady state (NESS). In  Figs.~\ref{fig:pcolor} (c) and (d), we present pseudocolor plots of the vorticity field, normalized by the maximum of $|\omega|$, for the parameters in Figs.~\ref{fig:pcolor} (a) and (b), respectively. These plots show clearly that the typical size of a single-phase domain decreases as activity-induced turbulence is enhanced by an increase in the value of $|\zeta|$.

We quantify active-turbulence-induced suppression of phase separation by plotting the coarsening length scale $\mathcal L(t)$ versus time $t$ in Fig.~\ref{fig:domain_growth}(a) for various values of $|\zeta|$; the plot for $\zeta = 0$ shows growth that is consistent with the Lifshitz-Slyozov form $\mathcal L(t) \sim t^{1/3}$ (dashed line~\footnote{In any DNS in a finite domain, $\mathcal L(t)$
approaches a finite value that is comparable to the linear size of the domain.}). As $t$ increases, $\mathcal L(t)$ saturates to a finite value for $|\zeta| > 0$, i.e., Eqs.~(\ref{eq:ch})-(\ref{eq:tensor}) lead to coarsening-arrest because of \textit{active turbulence}. In Fig.~\ref{fig:domain_growth}(b) we show how the mean coarsening-arrest scale $L_c = \langle\mathcal L(t)\rangle_t$ decreases as $|\zeta|$ increases (red curve); the attendant growth of the integral-scale Reynolds number $Re_{L_I}$ (blue curve) signals the enhancement of activity-induced turbulence.   

%$\mathcal L(t)$.  We recover the Lifshitz-Slyosov exponent $\mathcal L(t) \sim t^{1/3}$ for $\zeta = 0$ as shown in Fig.~\ref{fig:domain_growth}(a).
 % {\color{blue}In Fig.(?), we plot versus the simulation time, the total kinetic energy, $e(t)\equiv \int_{\mathcal{D}} \frac{1}{2}|\bm u|^{2} d\boldmath{x}$, and note that our DNSs have reached a steady state for all values $\zeta$. We also observe that for increasing values of $|\zeta|$, $e(t)$ increases; therefore activity injects energy into the system. }
%In Fig.~\ref{fig:domain_growth}(b) we plot versus $|\zeta|$ the time averaged length scale, $L_{c}\equiv\langle \mathcal{L}(t)\rangle_{t}$; we observe that with increasing $|\zeta|$, $L_{c}$ decreases implying an arrest in coarsening. While such results have been well known for externally driven binary fluid mixtures~\cite{perlekar2014spinodal,perlekar2017two, perlekar2019kinetic,berti2005turbulence}, our studies show similar coarsening arrest in a system that is self-driven. 

%The arrest in coarsening leads to increase in the interfacial area, in turn yields vortical structures and turbulence as shown in the pseudocolor plots of vorticity in Fig.~\ref{fig:pcolor}(c)-(d), similar to the bacterial tubbulence and turbulence in active nematics; here interface is defined as a diffusive region between dense and dilute phases. 

We now characterise the statistical properties of activity-induced turbulence in Eqs.~(\ref{eq:ch})-(\ref{eq:tensor}). 
% portray such turbulence by using  statistical quantities, such as kinetic energy spectrum, defined as follows:
%  \begin{eqnarray}
%      \mathcal E(k) = \left<\displaystyle \sum_{k\le k' <k+1} |\hat{\bm u}(\bm k', t)|^2\right>_t\,;
%  \end{eqnarray}
% here the average is taken in the statistically steady state. 
We begin with log-log plots of compensated energy and scalar-$\phi$ spectra, $k^{5/3}\mathcal E(k)$ and $k^{-7/6}\Phi(k)$, versus $k$, in Figs.~\ref{fig:spectra} (a) and (b), respectively, for various values of $|\zeta|$. These plots suggest that, as $\zeta$ increases,
the activity-induced turbulence in this systems leads to a nonequilibrium statistically steady state (NESS) with
an inertial range of scales in which the energy spectrum has a power-law form that is consistent with $\mathcal E(k) \sim k^{-5/3}$. We show below that this power-law spectrum arises because of an inverse cascade of energy. Its power-law form can then be surmised as in statistically steady homogeneous and isotropic 2D-fluid turbulence with an inverse energy cascade~\cite{boffetta2012two,pandit2017overview,kraichnan1967inertial,kraichnan1980two}. Note that this power-law region extends over nearly one-and-a-half decades of $k$ at the largest value $|\zeta|(= 1.5)$ that we consider. 

\begin{table}[h!]
    \small
		%\centering
		\begin{tabular}{| c | c | c | c | c | c |}
			\hline
			%\hspace{0.5cm}Run\hspace{0.5cm} & \hspace{0.5cm} $|\zeta|$ \hspace{0.5cm} & \hspace{0.55cm} $Re_{L_{I}}$ \hspace{0.55cm} & \hspace{0.5cm} $\alpha^{\prime}$ \hspace{0.65cm} & \hspace{0.65cm} $Pe$ \hspace{0.65cm} & \hspace{0.65cm} $We$ \hspace{0.65cm} \\
   Run&  $|\zeta|$  &  $Re_{L_{I}}$  &  $\alpha^{\prime}$  & $Pe$  & $We$  \\
			\hline
			R1& 0 & 0 & $-$ & 0 & 0\\
			\hline
			R2& 0.001 & $1.1\times10^{0}$ & $5.6\times10^{-2}$ &$2.8\times10^{-1}$ & $1.6\times10^{-4}$\\
			\hline
			R3& 0.01 & $3.2\times10^{0}$ & $7.9\times10^{-3}$ & $8.5\times10^{-1}$ & $2.3\times10^{-3}$ \\
			\hline
			R4& 0.03 & $1.3\times10^{1}$& $6.6\times10^{-3}$ & $3.5\times10^{0}$& $2\times10^{-2}$\\
			\hline
			R5& 0.05& $4.3\times10^{1}$ & $7.8\times10^{-3}$ & $1.2\times10^{1}$ & $1.1\times10^{-1}$\\
			\hline
			R6& 0.1 & $6.9\times10^{1}$ & $7.0\times10^{-3}$ & $1.9\times10^{1}$ & $2.3\times10^{-1}$\\
   			\hline
			R7& 0.5 & $1.1\times10^{2}$& $5.8\times10^{-3}$ & $3.2\times10^{1}$& $5.3\times10^{-1}$\\
			\hline
			R8& 1 & $1.2\times10^{2}$ & $6.8\times10^{-3}$ &$3.3\times10^{1}$ & $5.8\times10^{-1}$\\
			\hline
            R9& 1.5 & $1.3\times10^{2}$& $5.7\times10^{-3}$ & $3.4\times10^{1}$ & $6.1\times10^{-1}$\\
			\hline 
		\end{tabular}
		\caption{\label{tab:param} The values of various parameters in our DNS runs R1-R9.
		The following parameters are fixed in all these runs:
        $N = 1024, \; \text{grid size}\;\; dx = 2\pi/N, \; \epsilon = 3 dx, \; L = 2\pi, \; \text{Cn} = 3dx/L, \; M = \epsilon^2/2, \; \sigma = 1,\; \nu = 5\times 10^{-3}$, $\alpha = 0.01$. }
		\label{tab:param}
\end{table}
%%%%%%%%%%%%%%%%%%%%%%%%%%%%%%%%%%%%%%

We examine the energy-transfer mechanisms in the NESS of activity-induced turbulence in Eqs.~(\ref{eq:ch})-(\ref{eq:tensor}) 
by using the energy-budget equation~(\ref{eq:budget}) and evaluating the relevant scale-by-scale energy contributions~\cite{perlekar2019kinetic,alexakis2018cascades,verma2019energy}. In Figs.~\ref{fig:budget}(a) and (b) we present, for the illustrative values $|\zeta| = 0.001$ and (b) $|\zeta| = 1.5$, respectively, plots versus $k$ (log scale) of the inertial, friction, viscous, and active-stress contributions  $ T(k)\;({\rm{red}}),\; D_{\alpha}(k)\;({\rm{purple}}),\; D_{\nu}(k)\;({\rm{yellow}})$ and $S^{\phi}(k)\;({\rm{blue}}),$ which we have defined in Eq. (\ref{eq:fluxes}). Such plots indicate that, for low values of $|\zeta|$, the contributions of $T(k)$ and $D_{\alpha}(k)$ are negligible [Fig.~\ref{fig:budget}(a)], so, 
in the NESS with $\langle \partial_t \mathcal E(k, t)\rangle_t = 0$, dominant balance yields $ D_{\nu}(k) + S^{\phi}(k) = 0$. As $|\zeta|$ increases, both $T(k)$ and  $D_\alpha(k)$ increase in magnitude [Fig.~\ref{fig:budget}(b)], so a four-term balance is required in the NESS.

%Therefore, at higher activity, energy is injected by the active stress, and this energy is subsequently transferred to other wavenumbers through inertia. 
%Now we show the kinetic energy flux and active stress flux ~\cite{verma2019energy} for the $k_{th}$ wavenumber defined as 
%\begin{eqnarray}
%    \Pi(k) = -\int_{0}^{k'} T(k') dk' \label{eq:flux}
%\end{eqnarray}

To show that activity-induced turbulence exhibits a \textit{bona fide} inertial range we present plots for $|\zeta| = 0.001, 0.01, 0.1,$ and $1.5$  of the energy flux $\Pi(k)$ [Eq.~\ref{eq:fluxes}] versus $k$ (log scale) [Fig.~\ref{fig:flux}(a)].
We also present log-log plots versus $k$ of $|\Pi(k)|$ [Fig.~\ref{fig:flux}(b)] and $\Pi^{\phi}(k)$ [Fig.~\ref{fig:flux}(c)].
These plots show constant fluxes over at least one decade of the wavenumber $k$, so we have a well-defined inertial range, that has
a remarkable similarity with fluid turbulence~\cite{frisch1995turbulence}. The sign of $\Pi(k)$ in this range of scales indicates that activity-induced turbulence in Eqs.~(\ref{eq:ch})-(\ref{eq:tensor}) yields an \textit{inverse cascade} of energy that is reminiscent of a similar cascade in 2D statistically steady homogeneous and isotropic fluid turbulence~\cite{boffetta2012two,pandit2017overview,kraichnan1967inertial,kraichnan1980two}. By comparing the different plots in Fig.~\ref{fig:flux}(a), we see that this inverse cascade is suppressed as $|\zeta|$ decreases; this is reminiscent of similar cascade suppression in instability-driven 2D turbulence~\cite{van2022spontaneous}.

%$ T(k),\; D_{\alpha}(k),\; D_{\nu}(k),\; \rm{and}\; S^{\phi}(k)$ that are given in Eq.~(\ref{eq:fluxes}) and which we
%plot versus $k$ in Figs.~\ref{fig:budget}(a) and (b) for $|\zeta| = 0.001$ and $|\zeta| = 1.5$, respectively.

% \begin{eqnarray}
%     \partial_t \mathcal E(k, t) = T(k, t) + D_{\alpha}(k, t) + D_{\nu}(k, t) + S^{\phi}(k, t)\;\;\;\hspace{1cm}. \label{eq:budget}
% \end{eqnarray}
% The terms in the right hand side of Eq.~\ref{eq:budget} are
% \begin{eqnarray}
%     T(k, t) &=& - \mathfrak{R}\left[\displaystyle \sum_{k\leq |\bm k'|< k+1} [ \hat{\bm u}(-\bm k', t)\cdot \bm P(\bm k') \cdot \widehat{(\bm u \cdot \nabla \bm u)} (\bm k', t)]\right], \\
%     D_{\alpha}(k, t) &=& -2 \alpha \mathcal E(k), \\
%     D_{\nu}(k, t) &=& -2 \nu k^2 \mathcal E(k), \\
%     S^{\phi}(k, t) &=& \mathfrak{R}\left[\displaystyle \sum_{k\leq |\bm k'|< k+1} [ \hat{\bm u}(-\bm k', t)\cdot \bm P(\bm k') \cdot \widehat{(\nabla \cdot \bm \Sigma^A)} (\bm k', t)]\right],
%     \label{eq:contributions}
% \end{eqnarray}
% %\end{widetext}
% respectively, the energy transfer due to the inertial term, the energy disispation due to friction, the viscous disipation, and the energy transfer due to the active stress term. $P_{ij} = (\delta_{ij} - {k_ik_j}/{k^2})$ is the projection operator to ensure the incompressibility condition. 

\section{Conclusions}
\label{sec:conclusions}

We have uncovered activity-induced homogeneous and isotropic turbulence in the active Cahn-Hilliard-Navier-Stokes equations (CHNS), which provides a natural theoretical framework for our study, in which a single scalar order parameter $\phi$ [positive (negative) in regions where the microswimmer density is high (low)] is coupled with the velocity field $\bm u$. The activity of the microswimmers is governed by an activity parameter $\zeta$ that is positive for \textit{extensile} swimmers and negative for \textit{contractile} swimmers [see Eqs.~(\ref{eq:ch})-(\ref{eq:tensor})].  With extensile swimmers, this system undergoes complete phase separation, as in binary-fluid mixtures~\cite{tiribocchi2015active}. By carrying out extensive pseudospectral direct numerical simulations (DNSs) we have shown that this model develops an emergent nonequilibrium, but statistically steady, state (NESS) of active turbulence, for the case of contractile swimmers, if $\zeta$ is sufficiently large and negative. This turbulence arrests the phase separation into regions with positive and negative values of $\phi$, as in conventional fluid turbulence leads to the suppression of phase separation in a binary-fluid mixture~\cite{perlekar2014spinodal,perlekar2017two,perlekar2019kinetic}.

We have quantified this suppression by showing how the coarsening-arrest length $\mathcal{L}(t)$ scale does not grow indefinitely, with time $t$, but saturates at a finite value at large times. We have then characterised the statistical properties of this active-scalar turbulence by employing the energy spectrum $\mathcal E(k)$ and the fluxes $\Pi(k)$ and $\Pi^{\phi}(k)$. We have also obtained the spectrum of $\phi$, which is used in studies of phase separation. For sufficiently high Reynolds numbers, we have shown that the energy spectrum $\mathcal E(k)$ displays an inertial range, with a power-law dependence on the wavenumber $k$. We have demonstrated that, in this range, the flux $\Pi(k)$ assumes a nearly constant, negative value, which indicates that the system shows an inverse cascade of energy that is similar to its counterpart in 2D homogeneous and isotropic fluid turbulence, even though energy injection occurs over a wide range of wavenumbers in our active-CHNS model.

Our statistical characaterization of active-CHNS turbulence shows that it is fundamentally different from conventional 2D fluid turbulence~\cite{boffetta2012two,pandit2017overview}, forced 2D CHNS turbulence~\cite{perlekar2017two}, and other types of active-fluid turbulence, discussed, e.g., in Refs.~\cite{wensink2012meso, qi2022emergence,alert2022active, mukherjee2023intermittency,dunkel2013fluid, kaiser2014transport, joy2020friction,linkmann2019phase,linkmann2020condensate,bhattacharjee2022activity,aranson2022bacterial,kiran2023irreversibility}. For large values of $|\zeta|$, active-CHNS turbulence has some similarities with conventional 2D fluid turbulence and 2D forced CHNS turbulence, inasmuch as it shows an inverse-cascade region with $E(k) \sim k^{-5/3}$. The $|\zeta|$-dependent, small-$k$, power-law regime in $E(k)$ is qualitatively reminiscent of parameter-dependent small-$k$ power-law regimes in energy-spectra in some minimal models for bacterial turbulence~\cite{bratanov2015new,alert2022active,mukherjee2023intermittency, joy2020friction,kiran2023irreversibility,linkmann2019phase,linkmann2020condensate}. The scalar spectrum $\Phi(k)$ of active-CHNS turbulence shows a substantial power-law regime which is different from that in conventional forced 2D CHNS turbulence~\cite{perlekar2017two}. The fluxes and energy budgets in active-CHNS turbulence are also markedly different from their counterparts in  other types of 2D turbulence. Our study investigates the effects of active-CHNS turbulence that is distinct from the suppression of motility-induced phase separation in the active model H~\cite{tiribocchi2015active} with external noise.

 We hope, therefore, that our active-CHNS study will lead to investigations of experimental realisations of this system. Our results are of potential relevance to systems of contractile swimmers, e.g., \textit{Chlamydomonas reinhardtii}~\cite{yeomans2014introduction,fragkopoulos2021self} (\textit{C. reinhardtii}) and synthetic active colloids ~\cite{zottl2016emergent,howse2007self}. We look forward to the experimental verification of our results, especially in the former system, where it should be possible to control the activity by changing the oxygen concentration in low-light conditions.

%%%%%%%%%%%%%%%%%%%%%%%%%%%%%%%%%%%%
% \section*{Author Contributions} NBP, KVK, and RP planned the research and analysed the numerical data; NBP and KVK carried out the calculations and prepared the tables, figures, and the draft of the manuscript; NBP, KVK, and RP then revised the manuscript in detail and approved the final version.

% \section*{Conflicts of Interest}
% No conflicts of interests, financial or otherwise, are declared by the authors.

\section*{Acknowledgments}
We thank the Science and Engineering Research Board (SERB) and the National Supercomputing Mission (NSM), India for support, and the Supercomputer Education and Research Centre (IISc) for computational resources.
%%%%%%%%%%%%%%%%%%%%%%%%%%%%%%%%%%%%%%%%
\bibliography{Arxiv}

\end{document}